\documentclass[english,showpacs,preprint,superscriptaddress]{revtex4-1}
\usepackage{lmodern}
\usepackage{lmodern}
\usepackage[T1]{fontenc}
\usepackage{color}
\usepackage{amsmath}
\usepackage{amssymb}
\usepackage{graphicx}
\usepackage{esint}

\makeatletter

\usepackage{amsthm}\usepackage{latexsym}\usepackage{bm}\usepackage{amsfonts}

\usepackage{babel}

\usepackage{babel}
\usepackage{hyperref}

\makeatother

\usepackage{babel}
\begin{document}

\title{Multi-scale Full-orbit Analysis on Phase-space Behavior of Runaway
Electrons in Tokamak Fields with Synchrotron Radiation}

\author{Yulei Wang}

\affiliation{School of Nuclear Science and Technology and Department of Modern
Physics, University of Science and Technology of China, Hefei, Anhui
230026, China}

\affiliation{Key Laboratory of Geospace Environment, CAS, Hefei, Anhui 230026,
China}

\author{Hong Qin}

\affiliation{School of Nuclear Science and Technology and Department of Modern
Physics, University of Science and Technology of China, Hefei, Anhui
230026, China}

\affiliation{Plasma Physics Laboratory, Princeton University, Princeton, NJ 08543,
USA}

\author{Jian Liu}

\email{corresponding author: jliuphy@ustc.edu.cn}

\affiliation{School of Nuclear Science and Technology and Department of Modern
Physics, University of Science and Technology of China, Hefei, Anhui
230026, China}

\affiliation{Key Laboratory of Geospace Environment, CAS, Hefei, Anhui 230026,
China}
\begin{abstract}
In this paper, the secular full-orbit simulations of runaway electrons
with synchrotron radiation in tokamak fields are carried out using
a relativistic volume-preserving algorithm. Detailed phase-space behaviors
of runaway electrons are investigated in different dynamical timescales
spanning 11 orders. \textcolor{black}{In the small timescale}, i.e.,
the characteristic timescale imposed by Lorentz force, the severely
deformed helical trajectory of energetic runaway electron is witnessed.
A qualitative analysis of the neoclassical scattering, a kind of collisionless
pitch-angle scattering phenomena, is provided when considering the
coupling between the rotation of momentum vector and the background
magnetic field. In large timescale up to one second, it is found that
the initial condition of runaway electrons in phase space globally
influences the pitch-angle scattering, the momentum evolution, and
the loss-gain ratio of runaway energy evidently. However, the initial
value has little impact on the synchrotron energy limit. It is also
discovered that the parameters of tokamak device, such as the toroidal
magnetic field, the loop voltage, the safety factor profile, and the
major radius, can modify the synchrotron energy limit as well as the
strength of neoclassical scattering. \textcolor{black}{The maximum
runaway energy is also proved to be lower than the synchrotron limit
when the magnetic field ripple is considered.}
\end{abstract}
\maketitle

\section{Introduction\label{sec:Introduction}}

As a typical multi-scale process, the dynamics of runaway electrons
in tokamak has emerged as an important topic in the study of magnetic
confined fusion devices. During tokamak experiments, many operation
phases, such as fast shutdown, disruptions, and strong current drive,
are accompanied by the generation of runaway electrons \cite{Yoshino_Shutdown_RE,Jaspers_Disruption_RE,Helander_avalanch_2000,Helander_2002,Fulop_magneticThreshold4RE2009,Gill_REref1_2000,Jaspers_RErefs4_1993,Nygren_RErefs5_1997,Parks_RErefs6_1999,Rosenbluth_RErefs7_1997,Yoshino_REandTurbulenceDischarge2000,Tamai_Yoshino_REtermination_JT60U_2002,Lehnen_RMP_PRL_TEXTOR,Finken_RElosses_2007,Net_Fisch_RevModP1987}.
The collisional friction from the background plasma cannot prevent
the acceleration of these energetic electrons if the inductive loop
electric field is larger than a critical value \cite{Drercer_REorigins_1959,Connor_Relativistic_RE_1975}.
Through the acceleration by the electric field, the velocity of runaway
electrons can be sped up to nearly the light speed. Runaway electrons
carrying energies from 10 to 100 MeVs have been observed in different
experiments \cite{Bartels_RE_PFCs1994,Kawamura_PFS_RE1989,Bolt_REref0_1987,Jaspers_REref3_2001}.
Once hitting the plasma-facing components (PFCs), these energetic
electrons can damage the tokamak devices badly. Because of the strong
relativistic effect, the synchrotron radiation becomes an important
ingredient of runaway electron physics. For extremely energetic runaway
electrons, their synchrotron radiation loss could be strong enough
to balance out the acceleration by the loop electric field. The radiation
dissipation then provides runaway electrons an upper bound of energy,
i.e., the synchrotron energy limit \cite{Martin_Momentum_RE_1998,Martin_Energylimit_RE_1999,LiuJian_RE_Positron_2014,Guan_Qin_Sympletic_RE}.
The typical duration for a runaway electron with low energy (1keV-1MeV)
to reach the energy limit has the order of magnitude of one second
while the smallest timescale of Lorentz force is $10^{-11}\,\mathrm{s}$
\cite{LiuJian_RE_Positron_2014,CollisionlessScater_NF_Letter_2016},
which means the dynamical behavior of runaway electrons spans about
11 orders of magnitude in timescale. The multi-scale character poses
great difficulty to a satisfying physical treatment of runway dynamics. 

Through averaging out the gyro-motion, the gyro-center theory can
reduce the span of timescales by about three orders and is used widely
in dealing with runaway electron dynamics. Fruitful results of this
theory have been accomplished. Considering the gyro-center approximation
regardless of the toroidal geometry, one can transfer the full-orbit
dynamical equations of runaway electrons to a set of relaxation equations
which are much easier to solve theoretically and numerically \cite{LiuJian_RE_Positron_2014}.
By use of relaxation equations, the momentum evolution structure as
well as energy limit has been studied in detail under several kinds
of dissipations, such as collision, synchrotron radiation, and bremsstrahlung
radiation \cite{Martin_Momentum_RE_1998,Martin_Energylimit_RE_1999,Bakhtiari_Momentum_RE_Fb_2005}.
Meanwhile, the restriction effect of magnetic ripple on the maximum
energy has also been discussed in this way \cite{Martin_Energylimit_RE_1999}.
If involving the toroidal geometry, some extra geometry-related phenomena
arise, often dubbed neoclassical effects. The \textcolor{black}{Ware-pinch
effect sh}ows an inward drift of trapped orbit \cite{WarePinch_1970,Guan_Qin_Sympletic_RE,Fisch_Karney_1981_lowF_waves},
while the neoclassical drift provides an outward radial drift velocity
of transit runaway orbits \cite{Neoclassical_Drift_report}. Both
of these phenomena reflect the conservation of the toroidal canonical
angular momentum. Recently, gyro-center simulations have been equipped
with structure-preserving discrete methods and shown better long-term
numerical accuracy than traditional methods \cite{Guan_Qin_Sympletic_RE,LiuJian_RE_Positron_2014,Ruili_GC_canonical_2014}.

Unlike gyro-center theory, the full-orbit analysis can keep entire
physical information covering all timescales of runaway dynamics.
Especially, a recent full-orbit simulation on runaway electrons has
shown that the assumption of gyro-center theory no longer holds in
tokamak magnetic field if the runaway electrons are accelerated to
several tens of MeVs \cite{CollisionlessScater_NF_Letter_2016}. Because
of the high energy, the change of background magnetic field direction
encountered by runaway electrons is significant even within one gyro-period.
The violent change of magnetic field causes a full-orbit effect, the
collisionless neoclassical pitch-angle scattering, whic\textcolor{black}{h
arises}\textcolor{red}{{} }from the toroidal geometry and causes a violent
momentum exchange between parallel and perpendicular directions. It
also leads to a drift in momentum space and the significant run-up
of perpendicular momentum, which provides a new picture of runaway
momentum structure. The energy limit is also found to be higher when
the full-orbit effect is considered. Therefore, the full-orbit dynamical
analysis is vital to obtain reasonable descriptions on runways.

In this paper, we discuss the detailed full-orbit runaway dynamics
in views of both small ($10^{-11}$-$10^{-9}\,\mathrm{s}$) and large
($1$-$3\,\mathrm{s}$) timescales and analyze the influences of tokamak
design parameters on the long-term motion of runaway electrons.\textcolor{black}{{}
A throughout simulation of the multi-timescale behavior of runways
requires more than $10^{12}$ time steps, which is an astronomically
big number and cannot be properly implemented by traditional numerical
 methods. }To tackle the global accumulation of coherent errors for
such long-term simulation, we follow the method in Ref.\,\cite{CollisionlessScater_NF_Letter_2016}
and use a relativistic volume-preserving algorithm (VPA) \cite{Ruili_VPA_2015}.
As a geometric algorithm, the relativistic VPA possesses long-term
numerical accuracy and stability \cite{Guan_Qin_Sympletic_RE,HeYang_Spliting_2015,Jianyuan_Multi_sympectic_2013,LiuJian_RE_Positron_2014,Qin_Boris_2013,Qin_VariatianalSymlectic_2008,Ruili_GC_canonical_2014,Ruili_VPA_2015,HeYang_HamiltonTimeInt_VMs_2015,HeYang_HigherOrderVPA_2016,CSPIC_2016,XiaoJY_PIC_wave_2015,ExpNoncanonicalS_2015}.
The secular full-orbit dynamics of runaway electrons is obtained through
directly solving the Lorentz force equations. The synchrotron radiation
is included in the physical model, when the collisional force is ignored. 

The characteristic timescale imposed by magnetic force reflects the
smallest timescale of runaway dynamics, which can be defined as the
gyro-period
\begin{equation}
\mathcal{T}_{c}=\frac{2\pi\gamma\mathrm{m}_{0}}{\mathrm{e}B}\,,\label{eq:Tce}
\end{equation}
 where $\gamma$ is the Lorentz factor, $\mathrm{m}_{0}$ is the rest
mass of electron, $\mathrm{e}$ is the unit charge, and $B=\left|\mathbf{B}\right|$
denotes the strength of magnetic field. Although gyro-center theory
breaks down for energetic runaways in toroidal geometry, the gyro-period
can still be used as an available characteristic parameter for the
small timescale. This is because that the failure of the gyrocenter
condition is mainly due to the rapid change of the direction of the
magnetic field, while $B$ doesn't vary a lot during each gyro-period.
The practicability of $\mathcal{T}_{c}$ can also be analyzed in the
view of the rotation operator. We will show that in the gyro-period
timescale the trajectory of an energetic runaway electron is elongated
both toroidally and poloidally, and the corresponding $\mathcal{T}_{c}$
will increase to about one twentieth of the transit period.\textcolor{black}{{}
As a result, the local magnetic field witnessed by an energetic runaway
electron rotates rapidly, and the norm of magnetic rotation axial
vector, namely, $\left|\mathbf{\Omega}_{B}\right|=\left|\mathbf{b}\times\dot{\mathbf{b}}\right|$,
becomes comparable with $1/\mathcal{T}_{c}$, which leads to the collisionless
neoclassical pitch-angle scattering. A qualitative description of
the collisionless scattering is given through the coupling between
the rotations of momentum and magnetic vector. The momentum drift
caused by the long-term accumulation of collisionless scattering effect
is analyzed. To be specific, the perpendicular momentum of a runaway
electron increases in the direction of $-\mathbf{\Omega}_{B}$ which
is approximately in the direction of z-axis. }

The long-term evolution of momentum and energy are investigated for
runaways with different initial conditions in phase space.\textcolor{black}{{}
Four main characteristics of the momentum evolution structure are
discussed: }(a) the zero-point position of perpendicular momentum,
(b) the oscillation amplitude when reaching energy limit, (c) maximum
parallel momentum, and (d) maximum perpendicular momentum.\textcolor{black}{{}
Among these four characteristics, }(a) and (b) correspond to the fine
oscillating structures of runaway orbit, meanwhile (c) and (d) are
related closely to the synchrotron energy limit \cite{CollisionlessScater_NF_Letter_2016}.
It will be shown that the zero-point of perpendicular momentum and
the amplitude of oscillation are impacted significantly by the initial
pitch-angles. Larger initial perpendicular momentum will cause larger
zero-point position and stronger oscillation in small timescale. However,
the initial momentum samplings have little influence on the energy
limit. The impact of the initial configuration position on the long-term
momentum evolution is also \textcolor{black}{negligible.} For a deeper
insight, we define two quantities to describe the long-term integral
behavior of runaway energies, i.e., the energy loss-gain ratio and
the energy balance time. The energy loss-gain ratio is defined as
the ratio of the total energy loss through radiation to the energy
gained from the loop electric field. This ratio is influenced by the
initial runaway momentum significantly but is nearly independent of
the initial position. The evolutions of energy loss-gain ratio under
different initial phase space samplings have similar behaviors in
the vicinity of the energy limit. The energy balance time describes
the time required for a new-born runaway electron to run up to its
energy limit, which is approximately independent of its initial values
in the phase space. 

Finally, in order to \textcolor{black}{describe }tokamak experimental
research on runaways, the influences of tokamak parameters, including
the loop electric field, the background magnetic field, the major
radius, and the safety factor $q$, on both the energy limit and the
strength of neoclassical pitch-angle scattering are analyzed. Large
loop inductive electric field can impel runaways with high energy
in short time. On the other hand, the strength of magnetic field mainly
contributes to the neoclassical effects. Smaller magnetic field will
stall for the energy balance time but bump up the perpendicular momentum
more significantly. As a key parameter of tokamaks, the major radius
affects the energy limit and the balance time through changing the
power of radiation. Smaller major radius results in stronger radiation
and shorter balance time. The strength of neoclassical scattering
decreases slightly as the growth of major radius. The influence of
safety factor is also discussed. Involving several different effects,
the maximum energy, the balance time, and the maximum perpendicular
momentum roughly depend on the safety factor linearly. When $q$ is
small, the dependence of momentum oscillation on $q$ is more sensitive.
When $q$ is larger than 2, the amplitude of oscillation approaches
to a constant approximately. \textcolor{black}{Lastly, we also study
the effect of magnetic field ripple due to the finite number of toroidal
coils. The energy limit is proved to be lower than the synchrotron
limit when magnetic ripple exists, which is consistent with the theoretical
analysis in Ref.\,\cite{laurent_Rax_MagneticRipple1990}.}

This article is organized as follows. Section \ref{sec:Physical-Model}
gives an introduction of the physical model and the algorithm used
in the numerical research. In Sec.\,\ref{sec:Dynamics-in-Gyro-period},
the full-orbit behavior of a runaway electron is analyzed in the timescale
of $\mathcal{T}_{c}$. The long-term evolution behaviors of momentum
and energy are studied under different initial phase space samplings
in Sec.\,\ref{sec:MomentumStructure} and Sec.\,\ref{sec:Energy-integrated}
respectively. Section \ref{sec:Effects-of-Tokamak} focuses on how
the parameters of tokamak affect the energy limit and the neoclassical
pitch-angle scattering of runaway electrons. And Sec.\,\ref{sec:Summary}
concludes this paper and our future plans.

\section{Physical Model and numerical method\label{sec:Physical-Model}}

The first-principle physical model of runaway electron is the solution
of relativistic Lorentz force equations. The synchrotron radiation
is included as the dominate channel of runaway energy dissipation.
The collisional resistance is neglected because its effect is small
enough compared with the collisionless pitch-angle scattering \cite{CollisionlessScater_NF_Letter_2016}.
Consequently, we describe the runaway electrons by use of the following
equations,
\begin{equation}
\frac{\mathrm{\mathrm{d\mathbf{x}}}}{\mathrm{d}t}=\mathbf{v}\,,\label{eq:Lorentz_x}
\end{equation}
\begin{equation}
\frac{\mathrm{\mathrm{d\mathbf{p}}}}{\mathrm{d}t}=-\mathrm{e}\left(\mathbf{E}+\mathbf{v}\times\mathbf{B}\right)+\mathbf{F}_{R}\,,\label{eq:Lorentz_p}
\end{equation}
\begin{equation}
\mathbf{p}=\gamma\mathrm{m}_{0}\mathbf{v}\,,\label{eq:pgamma}
\end{equation}
 where $\mathbf{x}$, $\mathbf{p}$, $\mathbf{v}$ are respectively
position, momentum, and velocity of a runaway electron, $\mathbf{E}$
and $\mathbf{B}$ denote electric and magnetic field. The radiation
force is defined as \cite{CollisionlessScater_NF_Letter_2016}
\begin{equation}
\mathbf{F}_{R}=-P_{R}\frac{\mathbf{v}}{v^{2}}\,,\label{eq:Radiation_Grag}
\end{equation}
 where $P_{R}$ is the radiation power determined by \cite{Jackson_electrodynamics}
\begin{equation}
P_{R}=\frac{\mathrm{e}^{2}}{6\pi\mathrm{\mathrm{\epsilon}_{0}c}}\gamma^{6}\left[\left(\frac{\mathbf{a}}{\mathrm{c}}\right)^{2}-\left(\frac{\mathbf{v}}{\mathrm{c}}\times\frac{\mathbf{a}}{\mathrm{c}}\right)^{2}\right]\,.\label{eq:Pr}
\end{equation}
Here, $\mathrm{\epsilon}_{0}$ denotes the permittivity in vacuum,
$\mathrm{c}$ is the speed of light in vacuum, and $\mathbf{a}=\mathrm{d}\mathbf{v}/\mathrm{d}t$
is the acceleration vector. 

The full-orbit simulation of Eqs.\,\ref{eq:Lorentz_x}-\ref{eq:Pr}
is essentially a multi-scale numerical problem. To achieve the omni-timescale
dynamics of runaway electron, the minimum time resolution should be
less than $\mathcal{T}_{c}$, which is typically around $10^{-11}\,\mathrm{s}$.
However, since the timescale of acceleration process of runaway electrons
is one second, hundreds of billions of steps are needed in numerical
calculation. Traditional algorithms, such as the fourth-order Runge-Kutta
method, can only restrict one-step numerical error. So the global
coherent accumulation of numerical errors from such a large number
of simulation steps will go far beyond the tolerance of numerical
accuracy. To solve the numerical error problem for long-term simulations,
we deal with the problem by use of a relativistic volume-preserving
algorithm \cite{Ruili_VPA_2015}. The long-term numerical stability
and accuracy of the relativistic VPA has been verified. According
to the construction of relativistic VPA, the motion equations of runaway
electrons are discretized as 
\begin{equation}
\mathbf{a}_{k}=\frac{\mathbf{v}_{k}-\mathbf{v}_{k-1}}{\Delta t}\,,\label{eq:DesEq1}
\end{equation}
\begin{equation}
\mathbf{F}_{Rk}=\mathbf{F}_{R}\left(\mathbf{a}_{k},\mathbf{v}_{k}\right)\,,\label{eq:DesEq2}
\end{equation}
\begin{equation}
\mathbf{x}_{k+\frac{1}{2}}=\mathbf{x}_{k}+\frac{\Delta t}{2}\frac{\mathbf{p}_{k}}{\sqrt{m_{0}^{2}+\mathbf{p}_{k}^{2}/c^{2}}}\,,\label{eq:DesEq3}
\end{equation}
\begin{equation}
\mathbf{p}^{-}=\mathbf{p}_{k}-e\frac{\Delta t}{2}\mathbf{E}_{k+\frac{1}{2}}+\frac{\Delta t}{2}\mathbf{F}_{Rk}\,,\label{eq:DesEq4}
\end{equation}
\begin{equation}
\mathbf{p}^{+}=\mathrm{Cay}\left(\frac{-e\Delta t\hat{\mathbf{B}}_{k+1/2}}{2\sqrt{m_{0}^{2}+\mathbf{p^{-2}}/c^{2}}}\right)\mathbf{p}^{-}\,,\label{eq:DesEq5}
\end{equation}
\begin{equation}
\mathbf{p}_{k+1}=\mathbf{p}^{+}-e\frac{\Delta t}{2}\mathbf{E}_{k+\frac{1}{2}}+\frac{\Delta t}{2}\mathbf{F}_{Rk}\,,\label{eq:DesEq6}
\end{equation}
\begin{equation}
\mathbf{x}_{k}=\mathbf{x}_{k+\frac{1}{2}}+\frac{\Delta t}{2}\frac{\mathbf{p}_{k+1}}{\sqrt{m_{0}^{2}+\mathbf{p}_{k+1}^{2}/c^{2}}}\,,\label{eq:DesEq7}
\end{equation}
 where the subscript, $k$, denotes the $k$-th step, $\Delta t$
is the time interval, $\hat{\mathbf{B}}$ is defined as
\begin{equation}
\hat{\mathbf{B}}=\left(\begin{array}{ccc}
0 & B_{z} & -B_{y}\\
-B_{z} & 0 & B_{x}\\
B_{y} & -B_{x} & 0
\end{array}\right)\,,\label{eq:Bhat}
\end{equation}
and the symbol $\mathrm{Cay}$ denotes the Cayley transform \cite{Ruili_VPA_2015}.
The radiation force is treated as an effective electric field in the
discrete equations. In this paper, a typical configuration for tokamak
field is used, i.e.,
\begin{equation}
\mathbf{B}=-\frac{B_{0}R_{0}}{R}\mathbf{e}_{\xi}-\frac{B_{0}\sqrt{\left(R-R_{0}\right)^{2}+z^{2}}}{qR}\mathbf{e}_{\theta}\,,\label{eq:B}
\end{equation}
\begin{equation}
\mathbf{E}=E_{l}\frac{R_{0}}{R}\mathbf{e}_{\xi}\,.\label{eq:E}
\end{equation}
 Here we use the cylindrical coordinate system $\left(R,\xi,z\right)$.
In Eqs.\,\ref{eq:B} and \ref{eq:E}, $\mathbf{e}_{\xi}$ and $\mathbf{e}_{\theta}$
are respectively the toroidal and poloidal unit vectors, $R_{0}$
is the major radius, $q$ denotes safety factor, $E_{l}$ is the strength
of loop electric field, and $B_{0}$ is the magnitude of background
magnetic field. The time step of simulation is set as $\Delta t=1.9\times10^{-12}s$,
which is about 1\% of $\mathcal{T}_{c}$.

\section{Runaway dynamics in $\mathcal{T}_{c}$-timescale\label{sec:Dynamics-in-Gyro-period}}

In this section, we offer a straightforward full-orbit picture of
runaway electron dynamics in $\mathcal{T}_{c}$-timescale. Because
it has been proved that the gyro-center model breaks down for the
dynamics of energetic runaway electrons \cite{CollisionlessScater_NF_Letter_2016},
the motion of runaways in $\mathcal{T}_{c}$-timescale looks quite
different from the gyro-center picture. Here we set calculation parameters
based on a typical tokamak, that is $R_{0}=1.7\,\mathrm{m}$, $a=0.4\,\mathrm{m}$,
$q=2$, $B_{0}=2\,\mathrm{T}$, and $E_{l}=0.2\,\mathrm{V/m}$. The
initial position is chosen as $R=1.8\,\mathrm{m}$, $\xi=z=0$, and
the initial parallel and perpendicular momentums are set as $p_{\parallel0}=5\,\mathrm{m_{0}c}$
and $p_{\perp0}=1\,\mathrm{m_{0}c}$ respectively.

\begin{figure}
\includegraphics{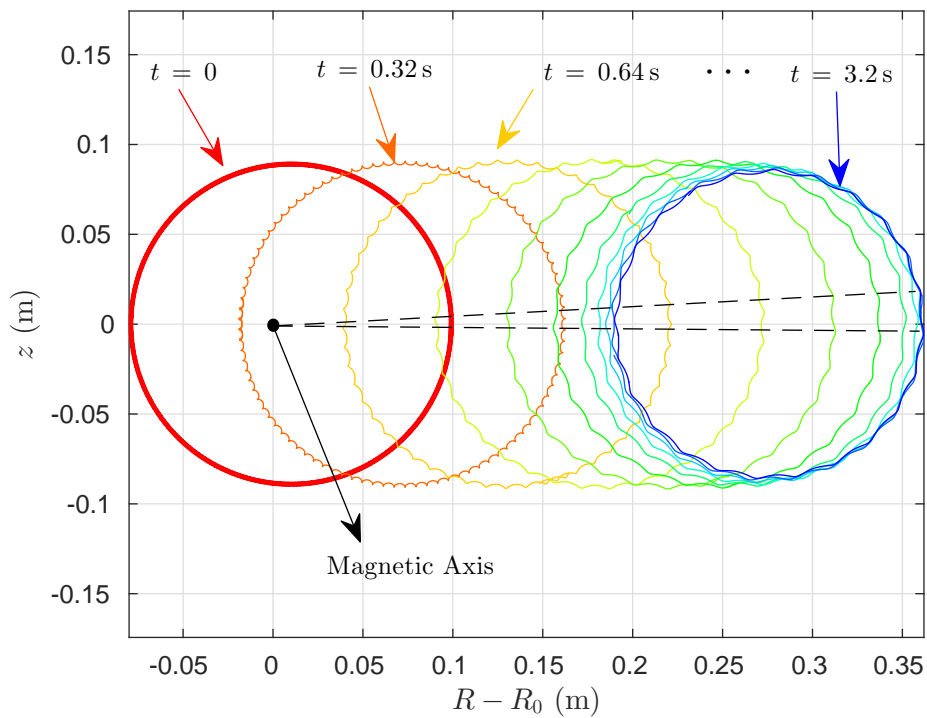}

\caption{Full-orbit snapshots of runaway orbit projected in the poloidal plane
at different moments. The configuration of field is determined by
$R_{0}=1.7\,\mathrm{m}$, $a=0.4\,\mathrm{m}$, $q=2$, $B_{0}=2\,\mathrm{T}$,
and $E_{l}=0.2\,\mathrm{V/m}$. The runaway electron is initially
sampled with momentum $p_{\parallel0}=5\,\mathrm{m_{0}c}$, $p_{\perp0}=1\,\mathrm{m_{0}c}$
at $R=1.8\,\mathrm{m}$, $\xi=z=0$. Besides the neoclassical radial
drift, the ripple structures are obviously exhibited.\textcolor{red}{{}
}\textcolor{black}{The poloidal angle spanned by one ripple at $t=3.2\,\mathrm{s}$
is marked by black dashed lines.}\label{fig:Drift}}
\end{figure}

Figure \ref{fig:Drift} depicts snapshots of poloidal projection of
runaway orbits at different moments. Besides the outward neoclassical
drift orbit similar to the results from the gyrocenter code \cite{Guan_Qin_Sympletic_RE,Neoclassical_Drift_report},
it can also be observed that there exist ripple structures superimposed
on each circle orbit. These fine ripple structures, which cannot be
recovered by gyro-center models, correspond to the runaway motion
in $\mathcal{T}_{c}$-timescale and become more obvious as time going.
Accompanying the increase of runaway energy, both the velocity and
$\mathcal{T}_{c}$ of runaway electrons grows. The runaway orbit during
each $\mathcal{T}_{c}$ is elongated both toroidally and poloidally.
As shown in Fig.\,\ref{fig:Drift}, the ripple structure becomes
more and more evident due to the enhance of the orbit elongation.
Because one circular orbit projected in the poloidal plane actually
corresponds to a transit period $\mathcal{T}_{tr}$ , the decrease
of the ripple structure number with energy increase implies the decrease
of $\mathcal{T}_{tr}/\mathcal{T}_{c}$ and hence the increase of $\mathcal{T}_{c}$
. For example, at $t=3.2\,\mathrm{s}$, there are only about 24 ripples
during one poloidal period, which means there are 24 $\mathcal{T}_{c}$s
in one transit period.

\begin{figure}
\includegraphics[scale=0.6]{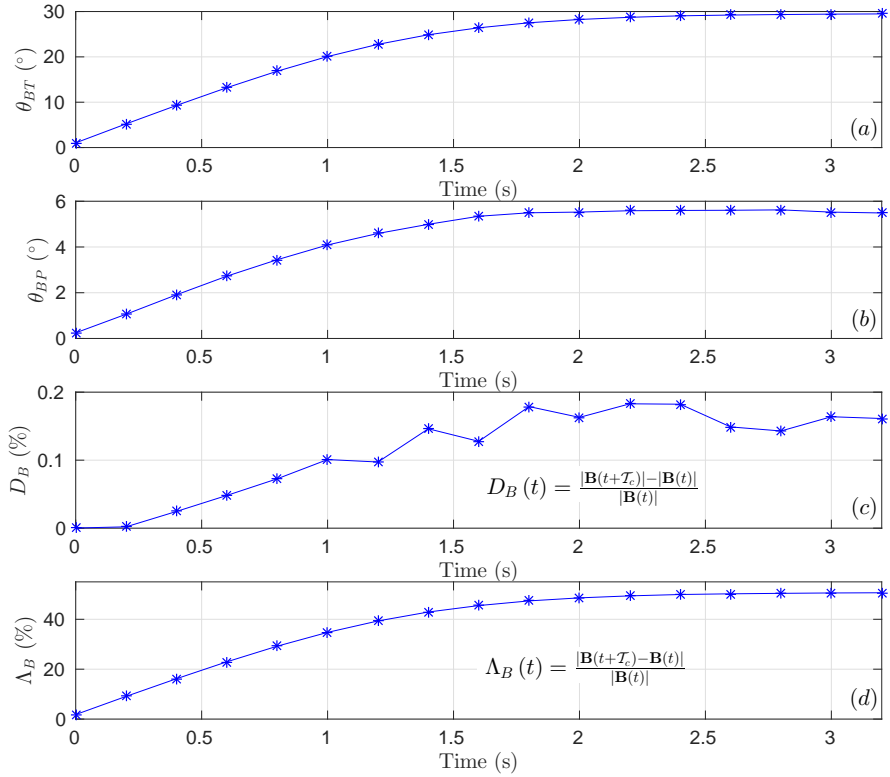}

\caption{The $\mathcal{T}_{c}$ timescale increment of $\mathbf{B}$ at differen\textcolor{black}{t
moments when the position of particle is near $z=0$. (a) depicts
the toroidal rotation angle $\theta_{BT}$, (b) gives the poloidal
rotation angle $\theta_{BP}$, (c) shows the relative change of magnetic
strength $D_{B}\left(t\right)$, and (d) depicts the relative increment
of magnetic vector $\Lambda_{B}\left(t\right)$. The maximum toroidal
rotation angle of the magnetic field in one gyro-period is $30^{\circ}$,
which is much larger than the corresponding poloidal rotation angle.
The increment of strength $B$ is too small to take effect as well.
Therefore, the change of magnetic field $\Lambda_{B}$ in $\mathcal{T}_{c}$
timescale is mainly due to the directional change of the magnetic
field torodcally.}\label{fig:Change-of-B}}
\end{figure}

The change of magnetic field witnessed by the runaway electron within
one gyro-period has a close relation to the deformation of runaway
orbit. The variance of the background magnetic field during $\mathcal{T}_{c}$
at different time is plotted in Fig.\,\ref{fig:Change-of-B}, indicated
by the toroidal rotation angle of magnetic field $\theta_{BT}$, the
poloidal rotation angle of magnetic field $\theta_{BP}$, the change
rate of magnetic strength $D_{B}$, and the relative increment of
magnetic vector $\Lambda_{B}$. \textcolor{black}{The definition $D_{B}\left(t\right)=\left(\left|\mathbf{B}(t+\mathcal{T}_{c})\right|-\left|\mathbf{B}(t)\right|\right)/\left|\mathbf{B}(t)\right|$
reflects relative change of magnetic strength within one gyro-period.
The relative increment of magnetic vector during one gyro-period is
defined as $\Lambda_{B}\left(t\right)=\left|\mathbf{B}(t+\mathcal{T}_{c})-\mathbf{B}(t)\right|/\left|\mathbf{B}(t)\right|$,
which includes the change of direction of the magnetic field. According
to Fig.\,\ref{fig:Change-of-B}c, during each gyro-period the increment
of the magnetic strength $D_{B}\left(t\right)$, which reflects the
small radial size of the ripple structure, is rather small compared
with $\Lambda_{B}\left(t\right)$. The poloidal and toroidal rotation
angles of magnetic field, $\theta_{BP}$ and $\theta_{BT}$, can be
approximately expressed by the poloidal and toroidal angles spanned
by a single ripple structure. The dashed lines in Fig.\,\ref{fig:Drift}
shows that the maximum poloidal angle spanned by a ripple is about
$6^{\circ}$, which is consistent with the result in Fig.\,\ref{fig:Change-of-B}b.
Compared with $\theta_{BP}$, $\theta_{BT}$ is significantly larger,
see Fig.\,\ref{fig:Change-of-B}a. The runaway electron runs about
$720^{\circ}$ in the toroidal direction and $360^{\circ}$ in the
poloidal direction within one transit period if $q=2$. Then at $t=3.2\,\mathrm{s}$,
24 ripples appearing in one transit period means that each ripple
structure covers about $30^{\circ}$ in the toroidal direction, which
is consistent with the maximum $\theta_{BT}$ in Fig.\,\ref{fig:Change-of-B}a.
Therefore, we can conclude that the violent change of magnetic field
in $\mathcal{T}_{c}$ timescale is dominated by its toroidal rotation. }

\begin{figure}
\includegraphics[scale=0.5]{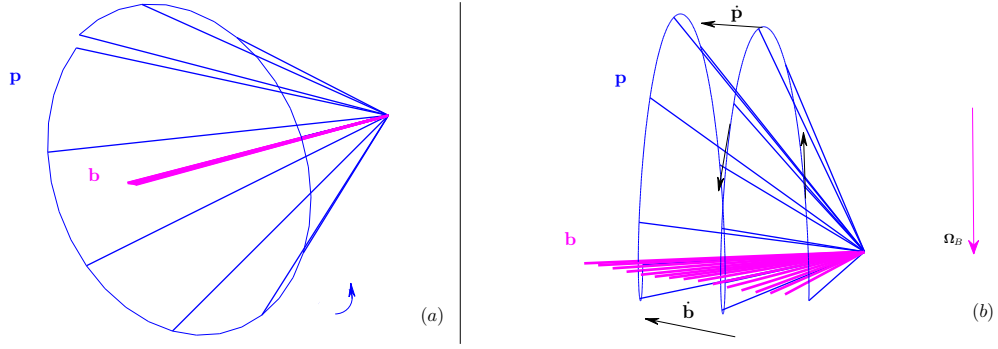}

\caption{Schematic diagram for the coupling between the rotations of momentum
and magnetic vector. The track of moementum vector spans a symmetric
cone around $\mathbf{b}$ if the magnetic field changes little in
the timescale of $\mathcal{T}_{c}$ (a). If the rotation rate of $\mathbf{b}$
is comparable to $1/\mathcal{T}_{c}$, the gyro-center assumption
breaks, and the track of momentum tiles upwards with respect to the
magnetic rotation plane (b). The rotation vector of $\mathbf{b}$
is marked by $\Omega_{B}$, $\dot{\mathbf{b}}$ denote\textcolor{black}{s
$\mathrm{d}\mathbf{b}/\mathrm{d}t$, and $\dot{\mathbf{p}}$ is $\mathrm{d}\mathbf{p}/\mathrm{d}t$.}
\label{fig:Schematic-diagram-BP}}
\end{figure}

\textcolor{black}{When analyzing the dynamics in the momentum space,
one can treat the effect of magnetic field as a rotation operation
due to the formation of Lorentz force }\cite{HeYang_Spliting_2015}.
The unit magnetic vector $\mathbf{b=B/}B$ determines the axis of
instantaneous momentum rotation, while the magnetic strength $B$
reflects the velocity of rotation as well as the value of $\mathcal{T}_{c}$.
If $\mathbf{b}$ is approximately a constant in $\mathcal{T}_{c}$,
the track of $\mathbf{p}$ spans a symmetric cone around $\mathbf{b}$,
see Fig.\,\ref{fig:Schematic-diagram-BP}a. \textcolor{black}{However,
if the characteristic variation time of $\mathbf{b}$ is comparable
with $\mathcal{T}_{c}$, the rotations of $\mathbf{b}$ and $\mathbf{p}$
are coupled, see Fig.\,\ref{fig:Schematic-diagram-BP}b, where $\mathbf{\Omega}_{B}$,
satisfying $\left|\mathbf{\Omega}_{B}\right|\sim1/\mathcal{T}_{c}$,
denotes the rotation axial vector of $\mathbf{b}$. Therefore, the
runaway's pitch-angle undergoes significant change even in the timescale
of $\mathcal{T}_{c}$. In Fig.\,\ref{fig:Schematic-diagram-BP}b,
$\dot{\mathbf{p}}$ and $\dot{\mathbf{b}}$ are marked using black
arrows, where the dot on physical quantities denotes total derivative
with respect to time. For runaway electrons carrying negative electric
charge, $\mathbf{p}$ always rotates counterclockwise with respect
to $\mathbf{b}$. The purple trails of $\mathbf{b}$ in Fig.\,\ref{fig:Schematic-diagram-BP}b
establish the rotation plane of $\mathbf{b}$. It is readily to see
that the value of $\dot{\mathbf{p}}\cdot\dot{\mathbf{b}}$ is positive
above the plane and negative under the plane. The coupling between
the rotations of $\mathbf{b}$ and $\mathbf{p}$ results in asymmetric
distribution of perpendicular momentum on two sides of magnetic rotation
plane, namely, the average of perpendicular momentum is larger when
$\dot{\mathbf{p}}\cdot\dot{\mathbf{b}}>0$.} Equivalently, the rotating
momentum can be regarded as being tilting towards the direction of
$-\mathbf{\Omega}_{B}$, i.e., the average $\left\langle \mathbf{p}_{\perp}\cdot\left(-\mathbf{e}_{rot}\right)\right\rangle $
always increases, where $\mathbf{e}_{rot}=\mathbf{\Omega}_{B}/\left|\mathbf{\Omega}_{B}\right|$
is the direction of rotation axial vector, and the bracket denotes
the averaging operation over $\mathcal{T}_{c}$. On the contrary,
a runaway positron carrying positive electric charge always rotates
around $\mathbf{b}$ clockwise. Consequently, the runaway positron's
momentum tilts towards the direction of $\mathbf{\Omega}_{B}$, namely
$\left\langle \mathbf{p}_{\perp}\cdot\mathbf{e}_{rot}\right\rangle $
increases. In the case of our simulation, the toroidal component of
magnetic field directs to $-\mathbf{e}_{\xi}$. Therefore, if neglecting
the poloidal component of $\mathbf{b}$, which is relative small,
we have $\mathbf{e}_{rot}=-\mathbf{e_{z}}$ approximately. Then the
average value of $z$-component of perpendicular momentum $\left\langle \mathbf{p}_{\perp}\cdot\mathbf{e}_{z}\right\rangle $
keeps growing. This effect is enhanced when the rotation of $\mathbf{b}$
becomes more rapid within one gyro-period. This theoretical analysis
agrees with the numerical results in Ref.\,\cite{CollisionlessScater_NF_Letter_2016}
and offers a direct description of the origin of collisionless pitch-angle
scattering.

\section{Secular Evolution of runaway Momentum \label{sec:MomentumStructure}}

Due to the collisionless neoclassical scattering, the temporal evolution
of runaway momentum shows strong oscillation in small timescale and
bumps up in large timescale. The structure of momentum evolution exhibits
complex multi-scale characteristics, which is different from the results
of gyro-center model. In this section, we aim to find the dependence
of the long-term momentum structure on the initial conditions of runaway
electrons in the phase space. The setup of tokamak parameters are
the same as those in Sec.\,\ref{sec:Dynamics-in-Gyro-period}.

\begin{figure}
\includegraphics{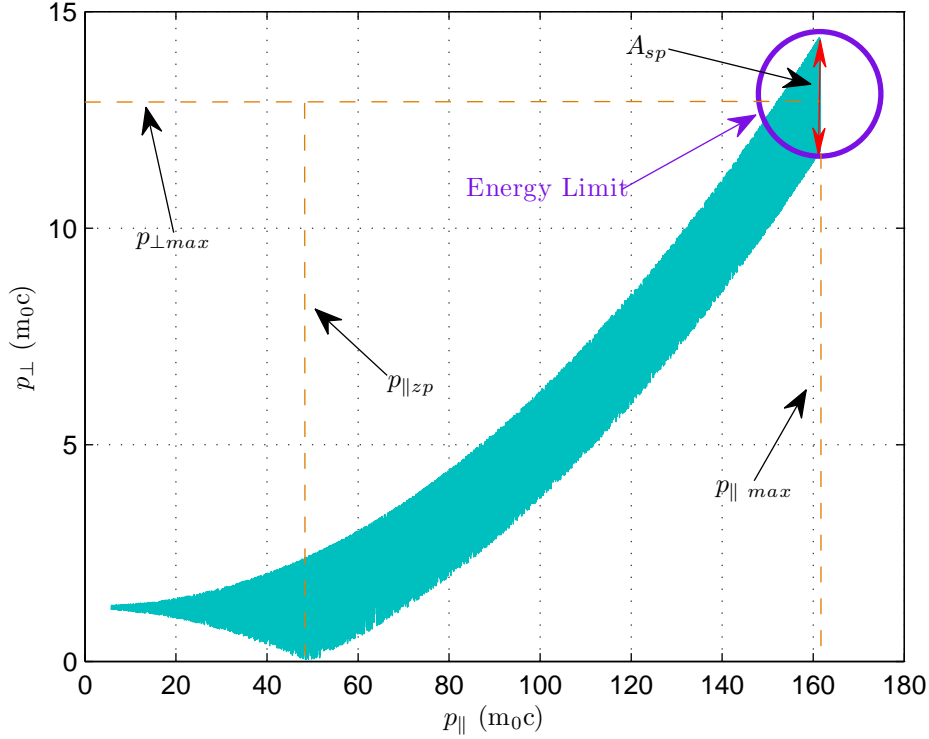}

\caption{A typical momentum evolution structure is plotted in the momentum
space, where the abscissa is the parallel momentum $p_{\parallel}$
and the ordinate is the perpendicular momentum $p_{\perp}$. The initial
momentum is set as $p_{\parallel0}=5\,\mathrm{m_{0}c}$ and $p_{\perp0}=1\,\mathrm{m_{0}c}$,
and the initial configuration position is $R=1.8\,\mathrm{m}$, $\xi=z=0$.
The parameters of the magnetic field are given by $R_{0}=1.7\,\mathrm{m}$,
$a=0.4\,\mathrm{m}$, $q=2$, $B_{0}=2\,\mathrm{T}$, and $E_{l}=0.2\,\mathrm{V/m}$.
The \textcolor{black}{momentum-space structure of runaway evolution
can be established by four principal parameters, that is the value
of parallel momentum at zero-point $p_{\parallel zp}$, }the limit
oscillation amplitude\textcolor{black}{{} $A_{sp}$, the maximum parallel
momentum $p_{\parallel max}$, and the average maximum perpendicular
momentum $p_{\perp max}$. }\label{fig:TypicalMomentum}}
\end{figure}

\textcolor{black}{A typical momentum evolution structure is plotted
in the momentum space, where the abscissa is the parallel momentum
$p_{\parallel}$ and the ordinate is the perpendicular momentum $p_{\perp}$,
see Fig.\,\ref{fig:TypicalMomentum}. The runaway electron starts
from $p_{\parallel0}=5\,\mathrm{m_{0}c}$ and $p_{\perp0}=1\,\mathrm{m_{0}c}$.
In the beginning , the oscillation amplitude increases significantly.
Then the perpendicular momentum touches $0$ at a zero-point $\left(p_{\parallel zp},\,0\right)$
due to the oscillation broadening. After passing the zero-point, the
global evolution of the momentum curve inclines to the ordinate, which
means the rapid increase of the perpendicular momentum. Finally, once
the loop electric field is balanced out by the radiation, the momentum
band ceases near }$p_{\parallel max}$\textcolor{black}{, with obvious
oscillation in }$p_{\perp}$\textcolor{black}{{} and negligible oscillation
in }$p_{\parallel}$\textcolor{black}{. The }$p_{\perp}$\textcolor{black}{{}
marked by the purple circle in Fig.\,\ref{fig:TypicalMomentum} corresponds
to the synchrotron energy limit. According to Fig.\,\ref{fig:TypicalMomentum},
the complete momentum-space structure of runaway evolution can be
basically established by four principal parameters, that is the zero-point
value of parallel momentum $p_{\parallel zp}$, }the oscillation amplitude
of perpendicular momentum near the energy limit\textcolor{black}{{}
$A_{sp}$, the maximum parallel momentum $p_{\parallel max}$, and
the average maximum perpendicular momentum $p_{\perp max}$. Among
them, $p_{\parallel zp}$ and $A_{sp}$ reflects the nature of momentum
oscillation and collisionless pitch-angle scattering, and $p_{\parallel max}$
and $p_{\perp max}$ provide the information of the energy limit \cite{CollisionlessScater_NF_Letter_2016}.}

\begin{figure}
\includegraphics[scale=0.9]{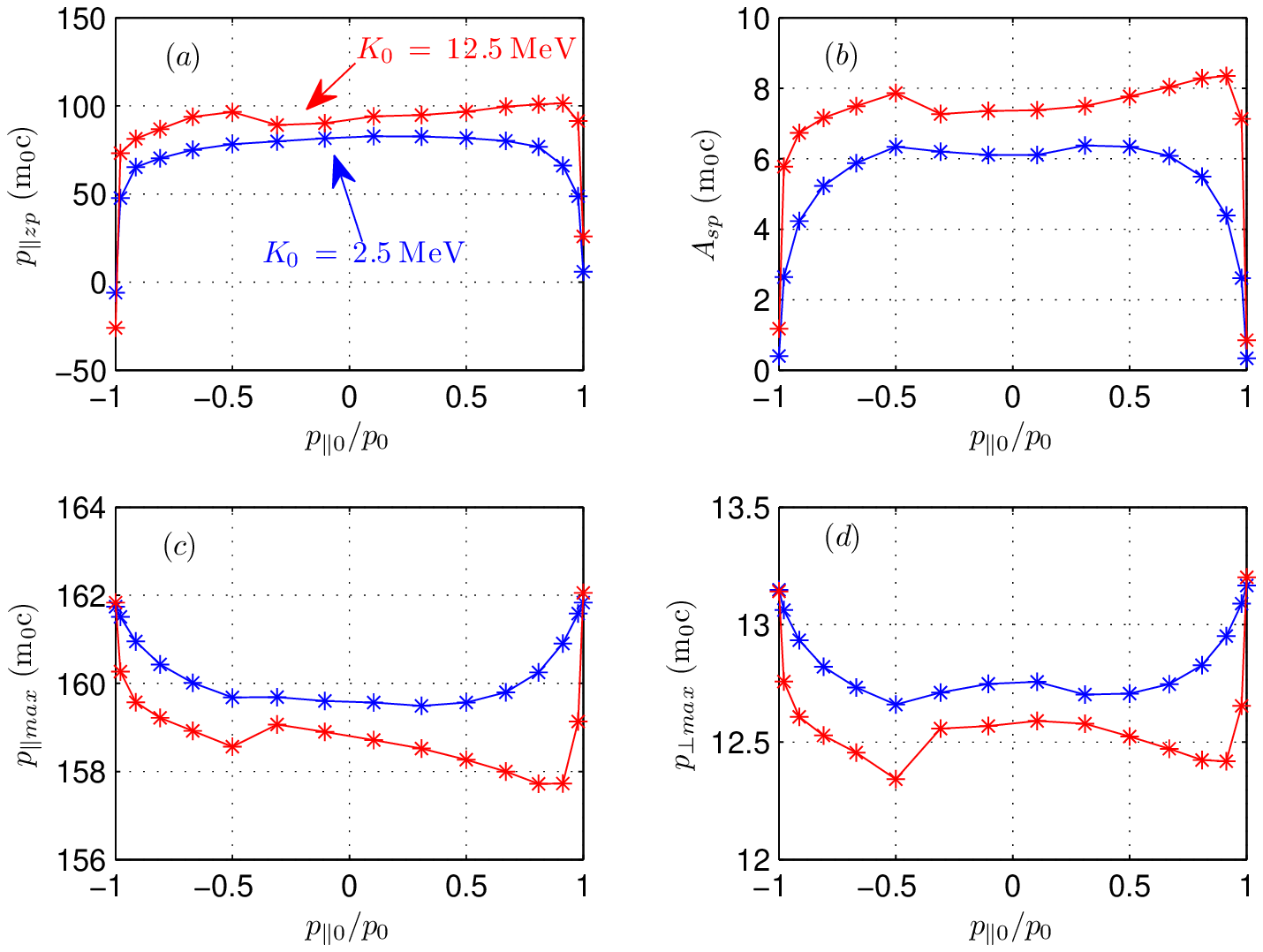}

\caption{Dependencies of key momentum structure parameters, i.e., (a) $p_{\parallel zp}$,
(b) $A_{sp}$, (c) $p_{\parallel max}$, and (d) $p_{\perp max}$,
on initial conditions in the momentum space in terms of the initial
kinetic energy and initial pitch-angle. The pitch angle is expressed
by $p_{\parallel0}/p_{0}$ ranging from -1 to 1, and the two initial
kinetic energy values $K_{0}=2.5\,\mathrm{MeVs}$ and $K_{0}=12.5\,\mathrm{MeVs}$
are chosen. The initial position is set to $R=1.8\,\mathrm{m}$, $\xi=z=0$.
The parameter of field is given by $R_{0}=1.7\,\mathrm{m}$, $a=0.4\,\mathrm{m}$,
$q=2$, $B_{0}=2\,\mathrm{T}$, and $E_{l}=0.2\,\mathrm{V/m}$. \label{fig:MomentumStruct}.}
\end{figure}

Different initial conditions of runaway electrons in the momentum
space mainly alter the position of zero-point and the amplitude of
oscillation, i.e., the properties of the neoclassical scattering,
but have little impact on the maximum momentum and energy limit. In
Fig.\,\ref{fig:MomentumStruct}, we calculate the dependencies of
key momentum structure parameters, i.e., (a) $p_{\parallel zp}$,
(b) $A_{sp}$, (c) $p_{\parallel max}$, and (d) $p_{\perp max}$,
on initial conditions in the momentum space, under different initial
kinetic energies and initial pitch-angles. The two sampling initial
kinetic energy values are chosen as $K_{0}=2.5\,\mathrm{MeVs}$ and
$K_{0}=12.5\,\mathrm{MeVs}$. The initial pitch-angles are uniformly
sampled from $0$ to $\pi$ in the range $\left[-1,1\right]$. The
negative value of $p_{\parallel0}$ means that the runaway electron
initially travels opposite to the electric acceleration direction,
i.e., the backward runaways \cite{Net_Fisch_RevModP1987,Karney_Fisch_1986}.
The effect of initial gyro-phase is not counted in considering the
gyro-symmetry in the low energy range. According to Fig.\,\ref{fig:MomentumStruct}a,
the \textcolor{black}{value of parallel momentum at zero-point} is
sensitive to the initial pitch-angle. It is readily to see that the\textcolor{black}{{}
value of parallel momentum at zero-point} tends to its initial value,
i.e., $p_{\parallel zp}\to p_{\parallel0}$, in the small pitch-angle
limit $p_{\parallel0}/p_{0}\to\pm1$. With the increase of $p_{\perp0}$,
or the decrease of $\left|p_{\parallel0}/p_{0}\right|$, the \textcolor{black}{value
of parallel momentum at zero-point} moves toward the positive direction
of $p_{\parallel}$. On the other hand, higher $K_{0}$ results in
larger $p_{\parallel zp}$. The oscillation amplitude of the momentum
structure is also closely related to initial momentum, see Fig.\,\ref{fig:MomentumStruct}b.
The amplitude $A_{sp}$ grows with the decrease of absolute value
of $p_{\parallel0}/p_{0}$ and the increase of initial energy. Especially,
if the initial pitch-angle is large enough, the oscillation amplitude
may catch up to one half of the maximum perpendicular momentum, see
Fig.\,\ref{fig:MomentumStruct}b and Fig.\,\ref{fig:MomentumStruct}d.
Conversely, as shown in Fig.\,\ref{fig:MomentumStruct}c and Fig.\,\ref{fig:MomentumStruct}d,
the initial momentum \textcolor{black}{has }relatively little effect
on $p_{\parallel max}$ and $p_{\perp max}$. The relative variations
resulted from different $p_{\parallel0}/p_{0}$ or $K_{0}$ are only
about 5\% or less.

\begin{figure}
\includegraphics[scale=0.9]{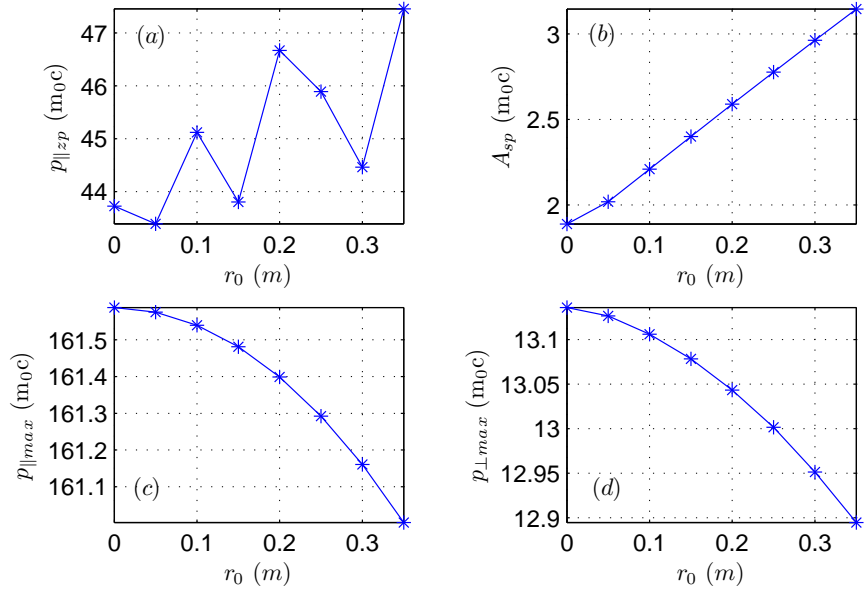}

\caption{Dependencies of key momentum structure parameters, i.e., (a) $p_{\parallel zp}$,
(b) $A_{sp}$, (c) $p_{\parallel max}$, and (d) $p_{\perp max}$,
on different initial samplings of radial position. The initial momentum
is set as $p_{\parallel0}=5\,\mathrm{m_{0}c}$ and $p_{\perp0}=1\,\mathrm{m_{0}c}$,
and the initial position is $R=1.8\,\mathrm{m}$, $\xi=z=0$. The
configuration of field is given by $R_{0}=1.7\,\mathrm{m}$, $a=0.4\,\mathrm{m}$,
$q=2$, $B_{0}=2\,\mathrm{T}$, and $E_{l}=0.2\,\mathrm{V/m}$. \label{fig:r0_effects} }
\end{figure}

The influence of initial positions in the configuration space on the
momentum evolution is \textcolor{black}{negligible}, as shown in Fig.\,\ref{fig:r0_effects}.
Since the initial position samplings possess approximate symmetry
in both toroidal and poloidal directions, the initial positions of
runaway electrons are sampled densely on radial positions. Here $r_{0}$
denotes the radial component of toroidal coordinates. The relative
variation of $p_{\parallel zp}$ is less than 6\% for different initial
radial positions in the range $r_{0}\in\left[0,\,0.35\right]$. For
larger $r_{0}$, both the electric field and the magnetic field witnessed
by a runaway electron are smaller when the transit orbit drifts away
from the magnetic axis. So less energy is delivered to runaway electrons
from electric field with the increase of $r_{0}$. We can see from
Figs.\,\ref{fig:r0_effects}c and \ref{fig:r0_effects}d that $p_{\parallel max}$
and $p_{\perp max}$ decrease slightly when $r_{0}$ gets larger,
but their relative variations are small. At the same time, because
smaller magnetic field implies stronger collisionless scattering \cite{CollisionlessScater_NF_Letter_2016},
the oscillation amplitude of momentum is proportional to $r_{0}$,
which is reflected in the plot of $A_{sp}$ in Fig.\,\ref{fig:r0_effects}b.

\section{Integral Attributes of Energy Evolution \label{sec:Energy-integrated}}

In this section, we focus on two important attributes of energy evolution,
namely, the energy loss-gain ratio and the energy balance time. The
energy loss-gain ratio of runaway electrons is defined as the ratio
of the energy dissipation through radiation to the energy gained from
the loop electric field. The energy balance time describes the time
required by a low-energy runaway electron from its birth to reaching
the energy limit. Unlike the energy limit, which can be analyzed through\textcolor{black}{{}
the stable point of a dynamical system,} the energy loss-gain ratio
and the energy balance time involves integral quantities over full
orbits.\textcolor{black}{{} Since it is too difficult to analytically
calculate the integrations for multi-timescale dynamics of runaway
electrons in tokamak fields, long-term numerical integ}ration turns
out to be the only practical way to achieve the accurate loss-gain
ratio and energy balance time. The parameters of tokamak we use in
this section are the same as those in Sec.\,\ref{sec:Dynamics-in-Gyro-period}.

\subsection{Energy loss-gain ratio}

\begin{figure}
\includegraphics[scale=0.45]{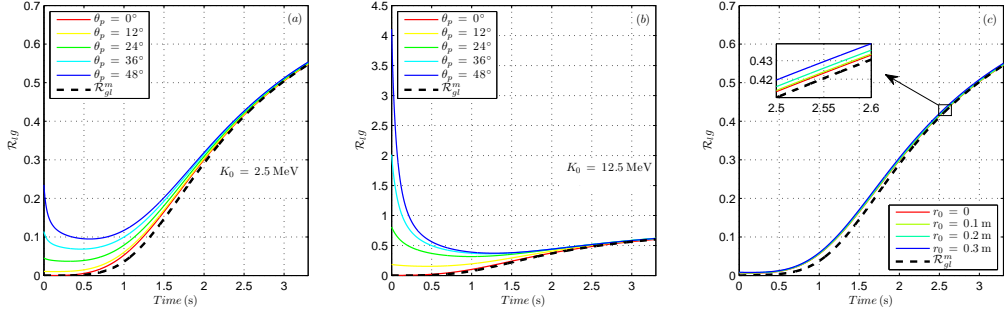}

\caption{The evolution of energy loss-gain ratio $\mathcal{R}_{lg}$ under
different initial energies, pitch-angles, and radial positions. Subfigures
(a) and (b) show the evolution of $\mathcal{R}_{lg}$ for $K_{0}=2.5\,\mathrm{MeV}$
and $12.5\,\mathrm{MeV}$ respectively. The initial pitch-angles in
(a) and (b) are chosen as $\theta_{p}=0^{\circ},\,12^{\circ},\,24^{\circ},\,36^{\circ},\,48^{\circ}$.
In (c), the initial radial position varies from $0$ to $0.3\,\mathrm{m}$
while the intial momentum is set as $p_{\parallel0}=5\,\mathrm{m_{0}c}$
and $p_{\perp0}=1\,\mathrm{m_{0}c}$. The influence of $r_{0}$ on
evolution of $\mathcal{R}_{lg}$ is negligible, and a zoomed-in window
showing the details of these curves is embeded in (c) to give a detailed
presentation. The dashed curves in all subfigures are reference lines
determined by $\mathcal{R}_{lg}^{m}\left(t\right)$, which are calculated
using the initial pitch angle $\theta_{p}=0^{\circ}$ in subfigures
(a) and (b) and using the initial radial position $r_{0}=0$ in (c).
\label{fig:Rlg} }
\end{figure}

The energy loss-gain ratio is defined as
\begin{equation}
\mathcal{R}_{lg}\left(t\right)=\frac{L_{s}\left(t\right)}{L_{s}\left(t\right)+G_{E}\left(t\right)}\,,\label{eq:Rlg}
\end{equation}
 where $L_{s}\left(t\right)=\left|\int_{0}^{t}\mathbf{F}_{R}\cdot\mathrm{d}\mathbf{x}\right|$
is the energy loss from the synchrotron radiation, and $G_{E}\left(t\right)=\int_{0}^{t}\mathbf{E}\cdot\mathrm{d}\mathbf{x}-L_{s}$
is the net runaway energy gained from electric field. After a runaway
electron reaches its energy limit, $G_{E}\left(t\right)$ becomes
a constant which is expressed using the symbol $G_{E}^{m}$. Therefore,
the energy loss-gain ratio has a simple form at energy limit, namely,
\begin{equation}
\mathcal{R}_{lg}^{m}\left(t\right)=\frac{L_{s}\left(t\right)}{L_{s}\left(t\right)+G_{E}^{m}}\,.\label{eq:Rlgm}
\end{equation}
 In Fig.\,\ref{fig:Rlg}, the evolution of $\mathcal{R}_{lg}$ is
plotted with different initial pitch-angles, kinetic energy, and radial
positions. From Fig.\,\ref{fig:Rlg}a and Fig.\,\ref{fig:Rlg}b,
we can see that the initial pitch-angle and energy mainly influence
$\mathcal{R}_{lg}$ at the early stage of acceleration. Larger pitch-angle
means smaller parallel velocity, which hence results in weaker electric
acceleration power and stronger radiation through perpendicular motion.
The increase of initial energy will enhance the radiation. Therefore,
$\mathcal{R}_{lg}$ grows together with the increase of $\theta_{p}$
and $K_{0}$. On the other hand, according to Fig.\,\ref{fig:Rlg}c,
the impact of initial radial position on the behavior of $\mathcal{R}_{lg}$
is very weak. All the curves in Fig.\,\ref{fig:Rlg} asymptotically
approach to the reference line of $\mathcal{R}_{lg}^{m}$ after $3\,\mathrm{s}$.
Therefore the initial position and momentum of runaways has little
effect on the energy loss-gain ratio after reaching the energy limit,
when about 55\% of the electric energy has been radiated.

\subsection{Energy balance time}

We now focus on how long it takes for a runaway electron to reach
its energy limit, i.e., the energy balance time $t_{blc}$. To calculate
$t_{blc}$, the start point and end point of acceleration process
should be determined. The end point is defined as the moment when
the loop electric field is balanced by the synchrotron radiation.
At this time, because of the neoclassical pitch-angle scattering,
physical quantities, such as momentum, electric acceleration power,
and radiation power, show strong oscillations in the gyro-period timescale.
The electric field is balanced out by the radiation loss only in the
sense of long-term average. Therefore, we define the average loss-gain
power ratio as
\begin{equation}
\eta_{RE}=\left\langle \frac{P_{R}}{P_{E}}\right\rangle _{trans}\,,\label{eq:etaRE}
\end{equation}
where $P_{R}=\left|\mathbf{F}_{R}\cdot\mathbf{v}\right|$ is the radiation
power, $P_{E}=\mathbf{E}\cdot\mathbf{v}$ is the electric acceleration
power and the bracket $\left\langle \cdots\right\rangle _{trans}$
means the average over a transit period. Then it is convenient to
define the end point of $t_{blc}$ as the moment when $\eta_{RE}=1$.
The end point of $t_{blc}$ can also be inferred from the relative
behavior of $\mathcal{R}_{lg}$ and $\mathcal{R}_{lg}^{m}$. Because
the evolutionary trend of $\mathcal{R}_{lg}$ becomes the same as
that of $\mathcal{R}_{lg}^{m}$ after reaching energy limit, the moment
when the curves of $\mathcal{R}_{lg}$ and $\mathcal{R}_{lg}^{m}$
begin to overlap also indicates the finish of acceleration.

The settlement of the start point of $t_{blc}$, however, is more
complex since runaway electrons are born with different phase-space
states and origins. There is even not a clear criterion for the emergence
of a single runaway electron because of its statistical essence. Fortunately,
it can be verified that different\textcolor{black}{{} initial energies
matters little to th}e energy balance time for runaways under several
MeVs. The typical electric field can accelerate a low-energy runaway
electron to several MeVs within 10\% $t_{blc}$ \cite{CollisionlessScater_NF_Letter_2016}.
In this paper, we set the start-up energy of a runaway electron a\textcolor{black}{s
$2.1\,\mathrm{MeV}$}. The arbitrariness of the setup of the start-up
energy has small impact on the value of $t_{blc}$ within the range
of several MeVs. According to Fig.\,\ref{fig:Rlg}, it can also be
observed that different initial samplings in phase space has little
effect on $t_{blc}$.

\begin{figure}
\includegraphics[scale=0.45]{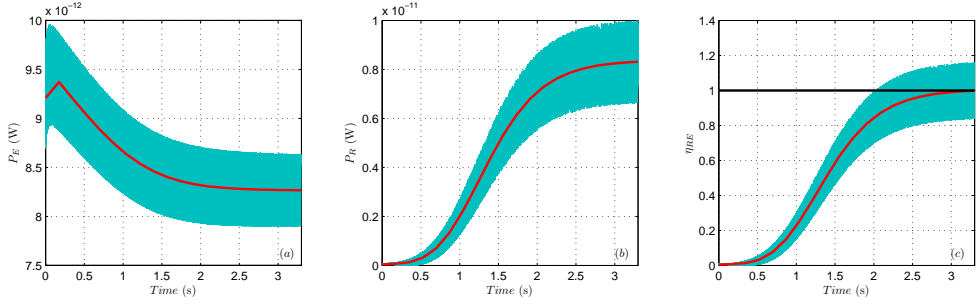}

\caption{Typical evolution curves of (a) electric acceleration power $P_{E}$,
(b) radiation power $P_{R}$, and (c) the average loss-gain power
ratio $\eta_{RE}$. The green bands denote the full-orbit evolution
curves including fine timescale oscilations. The red curves denote
the results after averaging over one transit period.\label{fig:TypicalEtaRE}}
\end{figure}

The typical evolutions of $P_{E}$, $P_{R}$, and $\eta_{RE}$ are
plotted in Fig.\,\ref{fig:TypicalEtaRE}. All of these curves show
strong oscillations. The transit-period average values are plotted
using the red curves. The power of electric acceleration increases
at the beginning because of the growth of runaway velocity. Once the
runaway speed is close enough to $\mathrm{c}$, the electric acceleration
pow\textcolor{black}{er is dominated by t}he strength of the loop
electric field. Since the electric field is inversely proportional
to radial position, $P_{E}$\textcolor{black}{{} decreases accompanied
by }the outward drift of runway transit orbits, see Fig.\,\ref{fig:TypicalEtaRE}a.
The radiation power monotonously increases with the runaway energy\textcolor{black}{{}
accompanied by} stronger oscillations, see Fig.\,\ref{fig:TypicalEtaRE}b.
The red curve in Fig.\,\ref{fig:TypicalEtaRE}c is a typical evolution
of $\eta_{RE}$, which reflects that the loss-gain rate power ratio
climbs steeply in the midterm of runaway acceleration. The energy
balance time is around $3\,\mathrm{s}$ in this case.

\section{Influences of tokamak Device Parameters\label{sec:Effects-of-Tokamak}}

Many experiments have confirmed the existence of runaway electrons
with energies ranging from $10$-$100\,\mathrm{MeVs}$ in tokamak
devices\cite{Gill_JET_REenergy_1993,Jarvis_JET_RE_1989,Wongrach_TEXTOR_disruption_2014}.\textcolor{black}{{}
To}\textcolor{red}{{} }\textcolor{black}{describe }and further understand
the experimental results, it is necessary to study the dependence
of runaway dynamical properties on the device parameters. In this
section several \textcolor{black}{characteristics o}f runaways, which
may be experimentally diagnosed directly or indirectly, such as the
maximum energy $E_{max}$, the energy balance time $t_{blc}$, the
oscillation amplitude of perpendicular momentum near the energy limit\textcolor{black}{{}
$A_{sp}$, and the average maximum perpendicular momentum $p_{\perp max}$,
are discussed.} The influences from three key tokamak device parameters
are considered, including the strength of tokamak field, the major
radius, and the safety fac\textcolor{black}{tor. The impact of magnetic
field ripple is also studied through full-orbit simulations.}

As vital design parameters, the intensities of equilibrium tokamak
fields are reflected in $E_{l}$ and $B_{0}$. The toroidal curvature
of fields is determined by the major radius $R_{0}$, while the poloidal
curvature is reflected in the safety factor $q$. All these parameters
influence the energy limit and collisionless pitch-angle scattering
by stepping in different aspects of runaway dynamics, such as the
acceleration, the synchrotron radiation, and the change rate of magnetic
field during each gyro-period. Larger $E_{l}$ leads to stronger acceleration,
while increasing $B_{0}$ results in the mitigation of collisionless
pitch-angle scattering. The increase of magnetic curvature corresponds
to the enhancement of radiation and toroidal effect. Considering the
loop electric field $\mathbf{E}$ and toroidal magnetic $\mathbf{B}$
decreases radially, the neoclassical drift velocity, approximately
given by $qE_{l}/B_{0}$ \cite{Guan_Qin_Sympletic_RE,Neoclassical_Drift_report},
also interferes the energy limit rule and the neoclassical scattering
process. The change of one single device parameter thus may affect
the runaway dynamics in several interactional mechanisms. On the other
hand, the magnetic field ripple can also impose stochastic instability
to runaway dynamics through the nonlinear resonance, which restricts
the maximum runaway energy below the synch\textcolor{black}{rotron
limit \cite{laurent_Rax_MagneticRipple1990}. I}n this section, the
initial conditions are sampled in the phase space the same as in Sec.\,\ref{sec:Dynamics-in-Gyro-period}.

\subsection{Influences of field strength }

\begin{figure}
\includegraphics{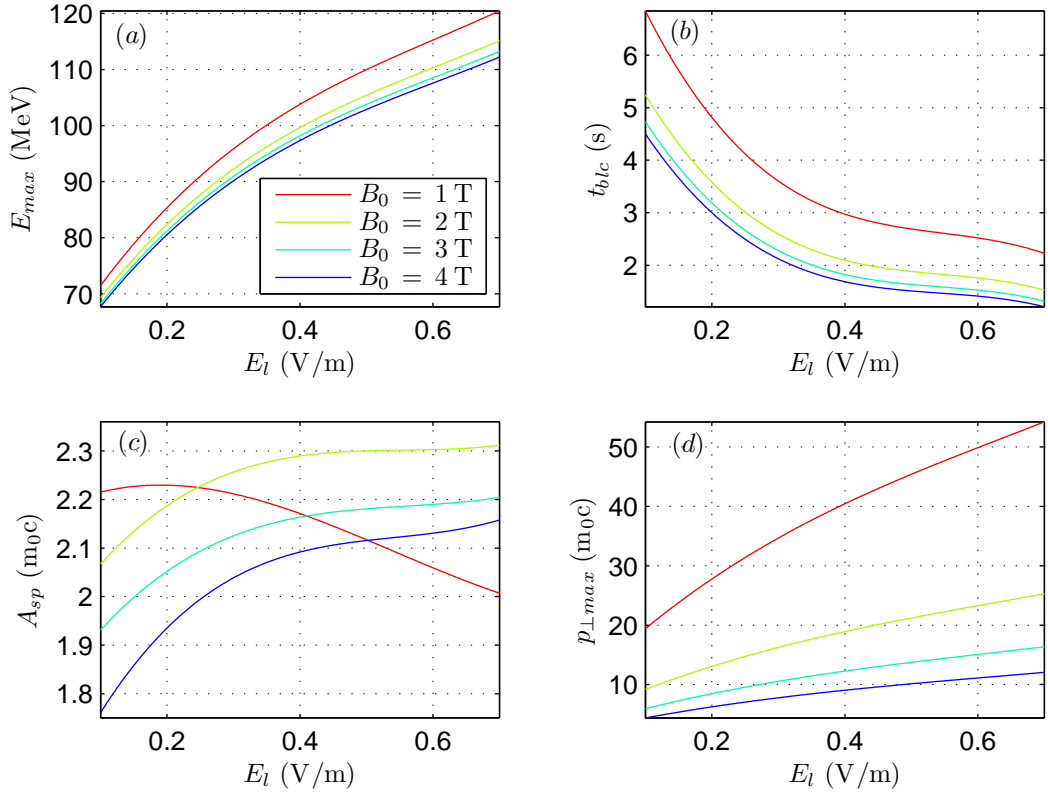}

\caption{The plots of (a) the maximum energy of a runaway electron, (b) the
energy balance time, (c) the oscillation amplitude at energy limit,
and (d) the maximun perpendicular momentum against loop electric field
with different magnetic field strengths. The initial condition is
set to $p_{\parallel0}=5\,\mathrm{m_{0}c}$, $p_{\perp0}=1\,\mathrm{m_{0}c}$,
$R=1.8\,\mathrm{m}$, and $\xi=z=0$. The major radius of tokamak
is $R_{0}=1.7\,\mathrm{m}$, and the safety factor is $q=2$.\label{fig:EB_Emax_Tblc_Azp_pperpmax}}
\end{figure}

Figure \ref{fig:EB_Emax_Tblc_Azp_pperpmax} summarizes the influences
from $E_{l}$ and $B_{0}$ on $E_{max}$, $t_{blc}$, $A_{sp}$, and
$p_{\perp max}$. Curves with different colors correspond to different
central magnetic field strength. The major radius is set as $1.7\,\mathrm{m}$
and the safety factor is $2$. In tokamak experiments, the loop electric
field has the strongest impact on the runaway energy limit. During
disruptions the energy of plasma is released through a strong inductive
loop electric field. According to Figs.\,\ref{fig:EB_Emax_Tblc_Azp_pperpmax}a
and \ref{fig:EB_Emax_Tblc_Azp_pperpmax}b, the runaways can reach
higher energy limits in shorter time when the loop electric field
increases, thus posing more severe threat in major disruption. 

On the other hand, the energy limit also depends on the magnetic field
significantly. From Fig.\,\ref{fig:EB_Emax_Tblc_Azp_pperpmax}a,
we can see that the energy limit is higher for smaller $B_{0}$. It
can be noticed that the increment of $E_{max}$ due to the drop of
$B_{0}$ is smaller than the result in Ref.\,\cite{CollisionlessScater_NF_Letter_2016}
which assumes a uniformly distributed electric field in the radial
direction. The difference is caused by the neoclassical drift and
radial distribution of electric field. For smaller $B_{0}$, as the
neoclassical drift is faster, the electric field witnessed by runaway
electrons decreases faster. Therefore, the energy limit is reduced
by several MeVs compared with that in the uniform electric field distribution.
This mechanism also results in longer balance time, see Fig.\,\ref{fig:EB_Emax_Tblc_Azp_pperpmax}b.

The strength of tokamak fields has small effects on the oscillation
amplitude in small timescale, see Fig.\,\ref{fig:EB_Emax_Tblc_Azp_pperpmax}c.
The largest relative change of oscillation amplitude caused by tokamak
field\textcolor{black}{{} is about 1}0\%. However, according to Fig.\,\ref{fig:EB_Emax_Tblc_Azp_pperpmax}d,
the ramp-up of perpendicular momentum in large timescale can be altered
significantly by adjusting the strength of tokamak field. The growth
rate of $p_{\perp max}$ versus loop electric field becomes much larger
for smaller magnetic field. This embodies stronger accumulation of
neoclassical pitch-angle scattering on perpendicular runaway momentum,
also more violent deviation from gyro-center model, for smaller magnetic
field and large electric field.

\subsection{Influences of Major Radius}

Next-generation tokamak devices possess larger major radius to achieve
higher operation parameters. For example, the major radius of ITER
is designed to be $6.2\,\mathrm{m}$. Larger tokamaks have better
confinement on fusion plasma as well as runaway electrons. More energy
may release through runaway currents during major disruptions. It
is obvious that runaway electron can gain more energy from the stronger
electric field. On the other hand, we will show that, even with the
same strength of electric field, the change of major radius $R_{0}$
will influence the energy limit rule and the neoclassical pitch-angle
scattering directly. In this part, the tokamak field is set as $E_{l}=0.2\,\mathrm{V/m}$
and $B_{0}=2\,\mathrm{T}$, and the safety factor is 2.

\begin{figure}
\includegraphics{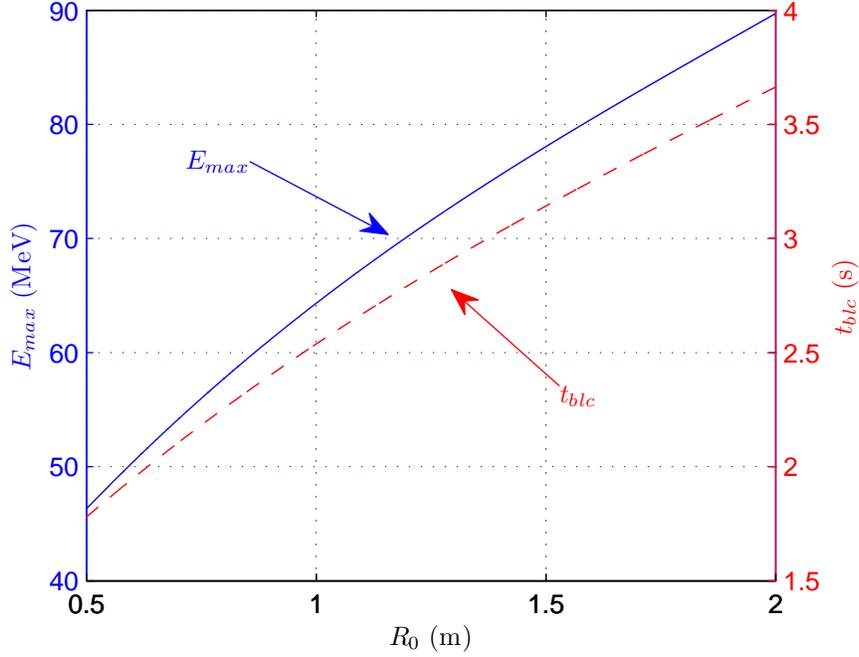}

\caption{The plots of energy limit $E_{max}$, denoted by the blue solid curve
and the left ordinate, and the energy balance time $t_{blc}$, denoted
by the red dashed curve and the right ordinate, against the major
radius. The initial condition of runaway electron in phase space is
sampled as $p_{\parallel0}=5\,\mathrm{m_{0}c}$, $p_{\perp0}=1\,\mathrm{m_{0}c}$,
$R=1.8\,\mathrm{m}$, and $\xi=z=0$. The tokamak field is set as
$E_{l}=0.2\,\mathrm{V/m}$ and $B_{0}=2\,\mathrm{T}$, and the safety
factor is chosen as $q=2$.\label{fig:Emaxtblc_R}}
\end{figure}

\begin{figure}
\includegraphics{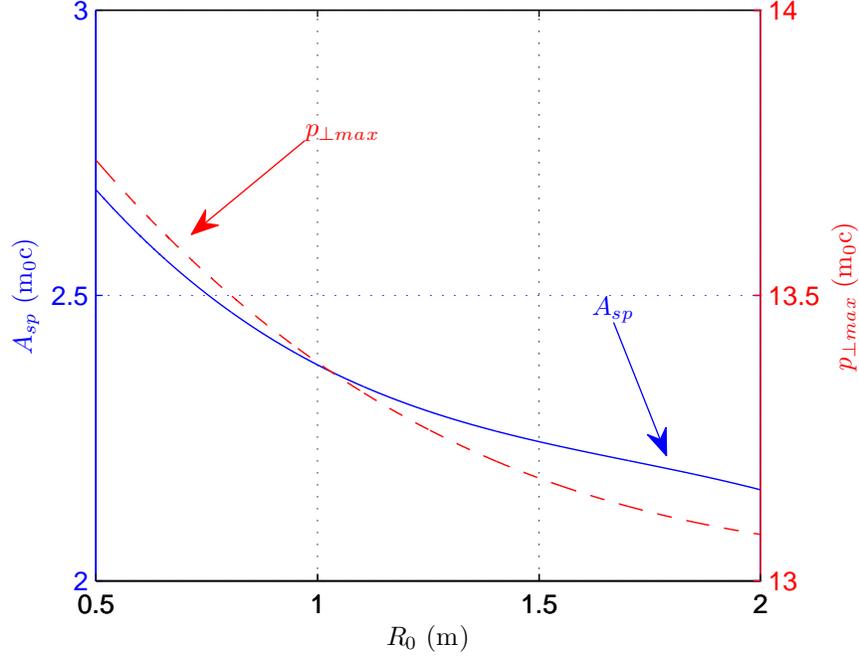}

\caption{The plots of magnitude of oscillation at energy limit $A_{zp}$, denoted
by the blue solid curve and the left ordinate, and the maximum perpendicular
momentum $p_{\perp max}$ , denoted by the red dashed curve and the
right ordinate, against the major radius. The initial condition of
runaway electron in phase space is sampled as $p_{\parallel0}=5\,\mathrm{m_{0}c}$,
$p_{\perp0}=1\,\mathrm{m_{0}c}$, $R=1.8\,\mathrm{m}$, and $\xi=z=0$.
The tokamak field is set as $E_{l}=0.2\,\mathrm{V/m}$ and $B_{0}=2\,\mathrm{T}$,
and the safety factor is $q=2$.\label{fig:R_Azp_Pperpmax}}
\end{figure}

Figure \ref{fig:Emaxtblc_R} plots the energy limit and the balance
time against different major radius. We can see that both the maximum
energy and balance time increase proportional to the major radius
$R_{0}$ approximately, because the synchrotron radiation closely
depends on the curvature of the runaway orbit \cite{Martin_Momentum_RE_1998}.
Smaller $R_{0}$ brings larger toroidal curvature and thus stronger
radiation power. The runaway electrons have stronger synchrotron dissipation
in small devices. As a result, their energy limit is lower and energy
balance time $t_{blc}$ is shorter.

The curvature of tokamak field reflects the significance of toroidal
geometry. So the major radius also affects the collisionless pitch-angle
scattering evidently. For devices with smaller major radius and larger
toroidal curvature, the same distance traveled in toroidal direction
brings more variation of the magnetic field direction. Consequently,
the assumption of gyro-center model breaks down easier in smaller
devices. As expected, the magnitude of oscillation at energy limit
$A_{zp}$ and the maximum perpendicular momentum $p_{\perp max}$
drop with the increase of $R_{0}$, see Fig.\,\ref{fig:R_Azp_Pperpmax}.

\subsection{Influences of safety factor}

The safety factor $q$ is another important parameter of tokamaks,
which reflects the geometric character of magnetic surface. Smaller
$q$ corresponds to stronger poloidal magnetic field and more poloidal
periods the magnetic line winding during each toroidal cycle. The
curvature of magnetic line is determined toroidally by the major radius
and poloidally by $q$. Therefore, dynamical processes related to
geometry configurations, such as synchrotron radiation and the neoclassical
pitch-angle scattering, will be influenced by $q$. \textcolor{black}{Larger
poloidal field also brings more difficult for electric field to accelerate
the runaway electrons toroidally. Meanwhile, the value of $q$ also
influence the neoclassical drift and thus the change of the local
strength of electric field. Generally speaking, the safety factor
$q$ has compound impacts on runaway dynamics, which makes its consequences
vague by analyzing any individual factor. During disruptions, large
portion of the poloidal magnetic field is induced by the runaway current.
So the $q$ profile in major disruption involves self-consistent evolution
of runway electrons.} In this part, we use the parameters $E_{l}=0.2\,\mathrm{V/m}$,
$B_{0}=2\,\mathrm{T}$, and $R_{0}=1.7\,\mathrm{m}$, while the value
of $q$ is sampled from 0.2 to 10.

\begin{figure}
\includegraphics{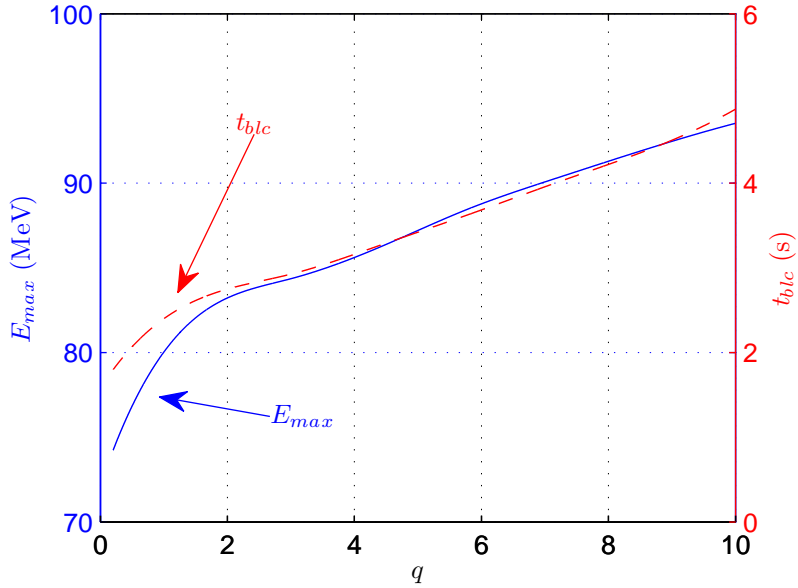}

\caption{The plots of energy limit $E_{max}$, denoted by the blue solid curve
and the left ordinate, and the energy balance time $t_{blc}$, denoted
by the red dashed curve and the right ordinate, against safety factor.
The initial condition of runaway electron in phase space is sampled
as $p_{\parallel0}=5\,\mathrm{m_{0}c}$, $p_{\perp0}=1\,\mathrm{m_{0}c}$,
$R=1.8\,\mathrm{m}$, and $\xi=z=0$. The tokamak field is set to
$E_{l}=0.2\,\mathrm{V/m}$ and $B_{0}=2\,\mathrm{T}$, and the major
radius $R_{0}=1.7\,\mathrm{m}$.\label{fig:q_Emax_Tblc}}
\end{figure}

Figure \ref{fig:q_Emax_Tblc} plots the energy limit and the energy
balance time against the safety factor $q$. Both $E_{max}$ and $t_{blc}$
increases as $q$ becomes larger. This phenomenon comes from two main
reasons. Firstly, for smaller $q$, the poloidal field is stronger.
Therefore, the toroidal acceleration of electric field is hindered.
Secondly, when the toroidal curvature determined by the major radius
keeps unchanged, smaller $q$ corresponds to larger poloidal curvature
and thus stronger synchrotron radiation. Even though the neoclassical
drift velocity is proportional to $q$ and the electric field decreases
faster for larger $q$ due to its radial distribution, the numerical
results in Fig.\,\ref{fig:q_Emax_Tblc} imply that the effect of
safety factor by modifying the neoclassical drift is weaker than the
above two effects.

\begin{figure}
\includegraphics{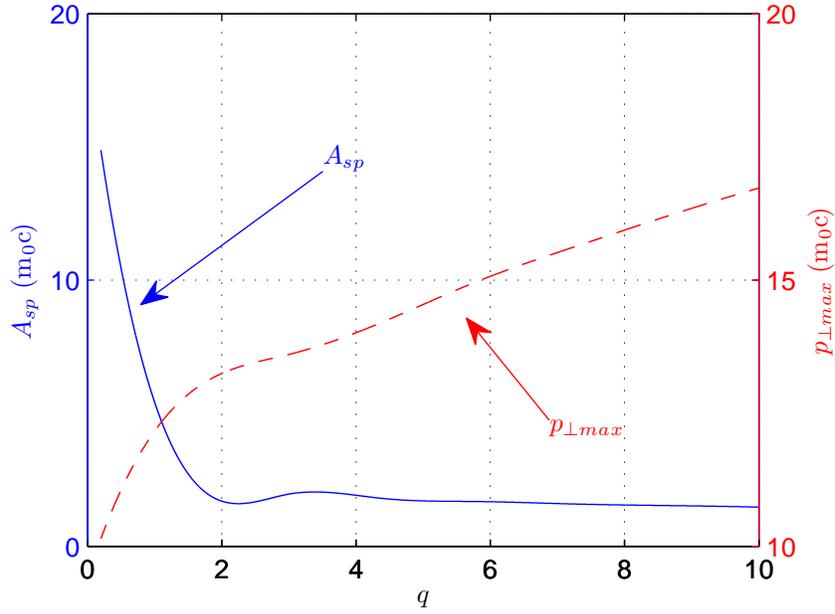}

\caption{The plots of magnitude of oscillation at energy limit $A_{zp}$, denoted
by the blue solid curve and the left ordinate, and the maximum perpendicular
momentum $p_{\perp max}$ , denoted by the red dashed curve and the
right ordinate, against safety factor. The initial condition of runaway
electron in phase space is sampled as $p_{\parallel0}=5\,\mathrm{m_{0}c}$,
$p_{\perp0}=1\,\mathrm{m_{0}c}$, $R=1.8\,\mathrm{m}$, and $\xi=z=0$.
The tokamak field is set to $E_{l}=0.2\,\mathrm{V/m}$ and $B_{0}=2\,\mathrm{T}$,
and the major radius is $R_{0}=1.7\,\mathrm{m}$.\label{fig:q_Asp_pperpmax}}
\end{figure}

The influences of $q$ on collisionless pitch-angle scattering and
maximum perpendicular momentum are plotted in Fig.\,\ref{fig:q_Asp_pperpmax}.
When the safety factor is less than 2, the amplitude of perpendicular
momentum oscillation drops evidently with the increase of $q$. Under
this condition, the neoclassical scattering is extremely strong, and
it is much easier for an energetic runaway electron to move cross
magnetic surfaces. The rotation of magnetic field witnessed by runaway
electrons becomes fast at the same time. The oscillation amplitude
even exceeds the maximum perpendicular momentum at energy limit when
$q$ is small enough. When $q$ is larger than 2, the geometric effects
mainly come from the toroidal field. The dependence of $A_{zp}$ on
$q$ is not notable. On the other hand, the plot of $p_{\perp max}$
shows the similar trend to that of $E_{max}$, which increases monotonously
with $q$.

\subsection{Influence of magnetic field ripples on energy limit}

In tokamaks, the toroidal magnetic fields are induced by toroidal
coils with finite coil number, which leads to magnetic field ripples
in experiments. According to L. Laurent and J. M. Rax's paper in 1990,
the stochastic instability caused by the nonlinear cyclotron resonances
with magnetic field ripples can transfer the parallel energy to perpendicular
direction and thus restrict runaway energy limit far below the synchrotron
energy limit \textcolor{black}{\cite{laurent_Rax_MagneticRipple1990}}.
In this subsection, utilizing full-orbit simulations, we study the
impacts of magnetic field ripples on the runaway energy limit. Besides
the equilibrium electromagnetic field given by Eqs\textcolor{black}{.\,\ref{eq:B}
and \ref{eq:E}}, there also exist the radial perturbation of magnetic
ripple \textcolor{black}{$\delta\mathbf{B}$,} expressed by
\begin{equation}
\delta\mathbf{B}=\delta B\mathbf{e}_{r}\,,\label{eq:Bripple}
\end{equation}
\begin{equation}
\delta B\left(r,\theta,\varphi\right)=\sum_{m=0,n=1}^{m=\infty,n=\infty}\delta B_{mn}\left(r\right)\mathrm{cos}\left(m\theta\right)\mathrm{cos}\left(nN\varphi\right)\,,\label{eq:MrippleInfty}
\end{equation}
\textcolor{black}{where, $r$, $\theta$, and $\varphi$ are three
components of the toroidal coordinates, $N$ is the number of toroidal
field coils, and $m$ and $n$ denote respectively the toroidal and
poloidal harmonics. Following the discussion in Ref.\,\cite{laurent_Rax_MagneticRipple1990},
we consider only the terms with $m=0,\,1$ in Eq. \ref{eq:MrippleInfty}.
The amplitude of perturbation magnetic field is given by the analytical
approximation for small $m$, namely,
\begin{equation}
\delta B_{0n}\left(r\right)=\delta B_{1n}\left(r\right)\approx\frac{B_{0}}{2}\left(1+\frac{qR_{0}}{R_{0}+b}\right)\mathrm{exp}\left[-Nn\left(\frac{b-r}{b+R_{0}}\right)\right]\,,\label{eq:DeltaB1n}
\end{equation}
where, $b$ is the radius of toroidal coils. Based on the Tore Supra
tokamak \cite{laurent_Rax_MagneticRipple1990}, we set the simulation
parameters as $N=18$, $B_{0}=1.8\,\mathrm{T}$, $E_{l}=0.1\,\mathrm{V/m}$,
$q=2$, $R_{0}=2.4\,\mathrm{m}$, $a=0.75\,\mathrm{m}$, and $b=1.3\,\mathrm{m}$.
The initial condition of the runaway electron is given by $p_{\parallel0}=5\,\mathrm{m_{0}c}$,
$p_{\perp0}=1\,\mathrm{m_{0}c}$, $r_{0}=0.1\,\mathrm{m}$, and $\theta_{0}=\varphi_{0}=0$.}

\begin{figure}
\includegraphics{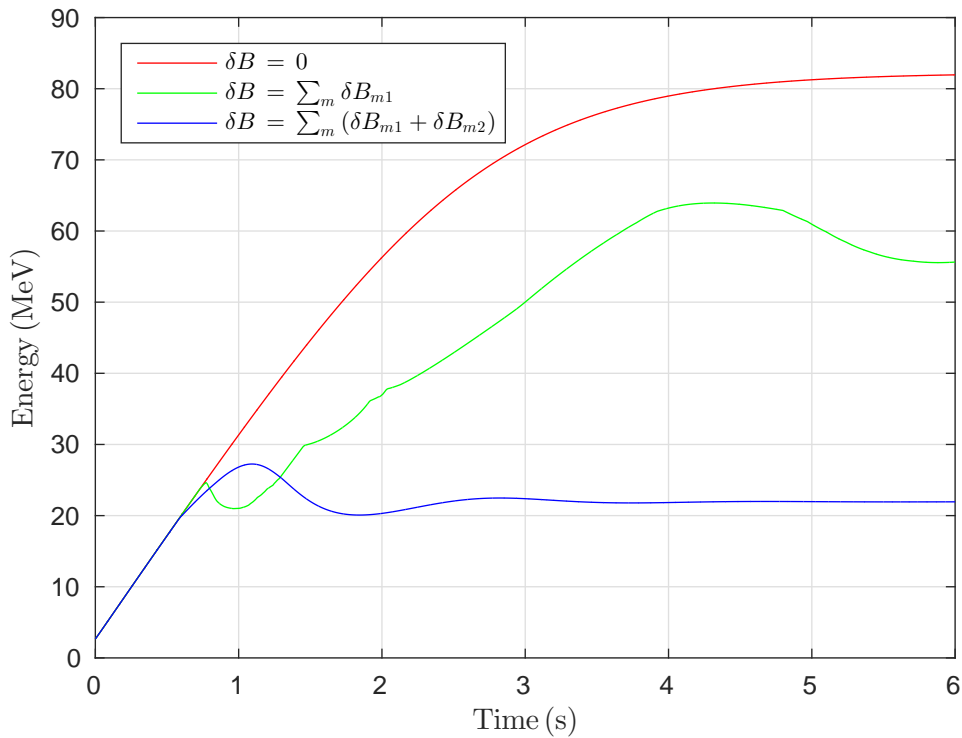}

\caption{The energy evolution of a runaway electron in tokamak fields with
different magnetic ripple perturbations. The red curve shows the energy
evolution without magnetic field ripple, the green one depicts the
result considering the $n=1$ components of $\delta B$, and the blue
one is the result considering both $n=1$ and $n=2$ harmonics of
$\delta B$. The summation over $m$ only covers two lowest components,
namely, $m=0,\,1$.\label{fig:MRipple_Energy}}
\end{figure}

Figure \ref{fig:MRipple_Energy} depicts the energy evolution of a
runaway electron affected by different harmonics of magnetic ripple.
Indicated by the red curve in Fig.\,\ref{fig:MRipple_Energy}, without
the magnetic field ripple, the synchrotron energy limit is about $80\,\mathrm{MeV}$.
When considering the $n=1$ components of ripple field, $\delta B_{01}$
and $\delta B_{11}$, the maximum runaway energy decreases to about
$60\,\mathrm{MeV}$ approximately, see the green curve in Fig.\,\ref{fig:MRipple_Energy}.
The restriction of the $n=2$ components of ripple field on runaway
energy is more significant. As shown by the blue curve, if we add
the components of $n=2$, namely, $\delta B_{02}$ and $\delta B_{12}$,
the energy limit is reduced to $22\,\mathrm{MeV}$. The results in
Fig.\,\ref{fig:MRipple_Energy} exhibit that the magnetic ripples
can limit the maximum runaway energy far below the synchrotron limit,
which is consistent with the results in Ref.\,\cite{laurent_Rax_MagneticRipple1990}.

\section{Conclusions\label{sec:Summary}}

In this paper, the multi-timescale runaway dynamics in tokamak field
is comprehensively exposed. The physical pictures in different timescales,
from $10^{-11}\,\mathrm{s}$ to $3\,\mathrm{s}$, have been fully
exhibited. The utilization of the relativistic volume-preserving algorithm
is vital to this study, because the long-term numerical accuracy and
stability of VPA ensures the accomplishment and correctness of the
secular numerical results. In the physical model, the toroidal configuration
of tokamak field and the synchrotron radiation are considered. Correspondingly,
the key role of geometric effects and coupling of multi-timescale
runaway dynamical processes are perfectly captured.

In small timescale imposed by Lorentz force, unlike the common wisdom,
the helical trajectory of energetic runaway electrons is elongated
both toroidally and poloidally so much that the collisionless neoclassical
scattering rises. A theoretical description of the neoclassical scattering
is provided through the coupling between the rotations of magnetic
field and momentum. The drift in momentum space is also analyzed based
on the rotation vector of magnetic field. The micro timescale dynamics
discussed in this paper has established a comprehensive picture of
runaway motion. More importantly, our results have shown that the
coupling between $\mathcal{T}_{c}$ and transit period plays an important
role for energetic runaways.

In large timescale up to several seconds, the long-term structure
of momentum evolution is portrayed by four characteristic quantities.
To find out the secular integral laws, we also studied the energy
gain-loss ratio and the energy balance time. The initial condition
is proved to have significant effects on small timescale momentum
oscillation but little influence on the long-term integral behaviors,
such as energy limit and energy balance time. \textcolor{black}{Meanwhile,
the dynamics of runaways can also be impacted by tokamak parameters.
The electromagnetic field, major radius, and safety factor have different
influences on both the energy limit and the neoclassical scattering
process through altering different aspects of runaway dynamics. It
is also proved that the existence of magnetic field ripple can reduce
the maximal runaway energy.}

Considering the complex influences from many different physical processes
in real tokamak discharges, we will study other factors on the energy
limit of runaway electrons, such as different instabilities and resonance
magnetic perturbations, in the future dynamical analysis of runaway
electrons. \textcolor{black}{Meanwhile, the observed runaway effects
in experiments are generally collective behaviors of large amounts
of runaway electrons. Therefore, the statistical treatment of runaway
}evolution with large samplings in the phase space will be carried
out to obtain macroscopic results, which is convenient for experimental
observation and verification.
\begin{acknowledgments}
This research is supported by National Magnetic Confinement Fusion
Energy Research Project (2015GB111003, 2014GB124005), National Natural
Science Foundation of China (NSFC-11575185, 11575186, 11305171), JSPS-NRF-NSFC
A3 Foresight Program (NSFC-11261140328), and the GeoAlgorithmic Plasma
Simulator (GAPS) Project. 
\end{acknowledgments}

\bibliographystyle{apsrev}
\bibliography{Refs_RunawayElectrons}

\end{document}